\documentclass[aps,aip,jmp,amsmath,twocolumn,nofootinbib]{revtex4-2}
\usepackage{bbold}
\usepackage{hyperref}
\def\[{\left\lbrack}
\def\]{\right\rbrack}

\def\({\left(}
\def\){\right)}
\newcommand{\be}{\begin{equation}}
\newcommand{\ee}{\end{equation}}
\newcommand{\ea}{\end{eqnarray}}
\newcommand{\ba}{\begin{eqnarray}}

\begin{document}

\title{Noncommutative Derivation of the Planck's Radiation Law}

\author{M. A. De Andrade}
\thanks{\noindent e-mail:marco@fat.uerj.br}

\author{L. G. Ferreira Filho}\thanks{e-mail:kph120@gmail.com} 

\author{C. Neves}
\thanks{\noindent e-mail:clifford@fat.uerj.br}
\affiliation{Departamento de Matem\'{a}tica, F\'\i sica e Computa\c{c}\~{a}o, \\
Faculdade de Tecnologia, \\ 
Universidade do Estado do Rio de Janeiro,\\
Rodovia Presidente Dutra, Km 298, P\'{o}lo
Industrial,\\
CEP 27537-000, Resende-RJ, Brazil.}
\begin{abstract}
\noindent The Planck's radiation law for the blackbody radiation spectrum was capable to explain the experimentally-measured blackbody spectrum. In order to get this result, Planck proposed his radiation law in a two-fold way: 1) by an \textit{ad hoc} modification of the assumed connection between energy and entropy for thermal radiation; 2) by assuming that the calculation of the entropy of an oscillator in thermal equilibrium with radiation is carried out by discrete units of energy. As a consequence, the energy quantization, linear in frequency, was thus introduced into physics. However, the energy quantization of the simple harmonic oscillator was originally postulated by Planck in an incomplete way, i.e., the ground state energy was not assumed to be null. Of course, this issue has been solved in different ways over time. Despite of this, we propose an alternative way to fix this issue by describing harmonic oscillators at noncommutative(NC) framework, where the non-null ground state energy naturally arises as a NC contribution. With this approach, the Planck's quantum theory is updated and, consequently, becomes compatible with the quantum mechanics inaugurated in 1925.
\end{abstract}

\maketitle

\setlength{\baselineskip} {20 pt}

\section{Introduction}
\label{int}

Usually, the quantum concept -- energy quanta linear in frequency -- appears in
modern physics textbooks\citep{books} through a theoretical derivation of the blackbody energy
spectrum for thermal radiation, demonstrating the failure
of classical theory and in proposing the formulation of
a new mechanics, i.e., quantum mechanics. Of course, quantum mechanics has been developed far beyond the
original modifications of classical theory demanded for
the derivation of the Planck's radiation law.

However, some interesting papers\cite{Boyer1,Boyer2,Boyer3,Boyer4} present derivations of the experimentally-measured blackbody energy spectrum and the non-null ground state energy without quantum assumptions (discrete energy), but requiring in addition to the usual ideas of classical theory some classical assumptions, for example: Lorentz-invariant electromagnetic radiation at the absolute zero of temperature, or exploring some connections between classical and quantum theories for the harmonic oscillator or applying purely thermodynamic theory to the classical simple harmonic oscillator. 

On the other hand, the desire to describe quantum mechanics on phase space instead Hilbert space is old as quantum mechanics itself\cite{Wigner,Groenewold,Moyal} and others have contributed to the subject\cite{Groot,Bayen1,Bayen2,lesche1,lesche2}. 
In this scenario, we propose to obtain the Planck's radiation law for the full blackbody energy spectrum. This is accomplished by using the $\star$-product instead scalar product among phase space variables. Due to this, the ground state energy $\hbar\omega/2$ arises as a NC contribution and, as a consequence, the Planck's postulate for simple harmonic oscillators and the Planck's radiation law are updated to also embrace this ground state energy.

\section{Updating Planck's radiation law through Noncommutativity}
\label{sec1}
The noncommutativity\cite{Groenewold,Moyal} will be introduced into the Rayleigh-Jeans theorem\cite{books} and, consequently, the seminal issue in the Planck's quantum theory\cite{Hermann} will be solved. In order to get this, the scalar product among the phase space variables will be changed to the $\star$-product\cite{Groenewold,Moyal}, given by
\be 
\label{NovoProduto}
f(\xi) \star g(\xi) = f(\xi) ~exp\left[\frac {i\hbar}{N}~\left(\frac{\overleftarrow{\partial}}{\partial Q_i}~\frac{\overrightarrow{\partial}}{\partial P_i}-\frac{\overleftarrow{\partial}}{\partial P_i}~\frac{\overrightarrow{\partial}}{\partial Q_i}\right) \right]~g(\xi),
\ee
where $\frac{\overleftarrow{\partial}}{\partial Q_i}$ e $\frac{\overrightarrow{\partial}}{\partial P_i^\prime}$ mean, respectively, derivative to the left and to the right, and $N$ is an arbitrary constant. Imposing that $\hbar/N<<1$, Eq.~(\ref{NovoProduto}) renders to
\be 
\label{NovoProduto1}
f(\xi) \star g(\xi) = f(\xi)\left[1 + \frac {i\hbar}{N}~\left(\frac{\overleftarrow{\partial}}{\partial Q_i}~\frac{\overrightarrow{\partial}}{\partial P_i}-\frac{\overleftarrow{\partial}}{\partial P_i}~\frac{\overrightarrow{\partial}}{\partial Q_i}\right) \right] g(\xi). 
\ee
 The energy of a harmonic oscillator in a noncommutative framework is
\ba  
\label{0001}
W&=&\frac 12\left(\textbf{p}\star\textbf{p} +\omega^2\textbf{x} \star \textbf{x}\right),\nonumber\\
&=&\frac {\left(\textbf{p}-\imath\omega\textbf{x}\right)}{\sqrt{2}} \star \frac {\left(\textbf{p}+\imath\omega\textbf{x}\right)}{\sqrt{2}},\nonumber\\
&=& \frac 12\left(\textbf{p}^2+\omega^2\textbf{x}^2\right) +\frac{\hbar\omega}{N},
\ea 
where the usual scalar product among the phase space variables was replaced by the $\star$-product. This allows us to update the Planck's postulate to 
\be 
\label{0010}
W_n=\left(n + \frac 12\right)\hbar\omega,
\ee
with $N=2$. Note that for $n=0$ there is a non-null ground state energy $(\hbar\omega/2)$ due to the noncommutativity. In this scenario, the Planck quantum theory reproduces the result obtained in the modern quantum mechanics establish in 1925\cite{MQ1,MQ2,MQ3,MQ4,MQ5,MQ5,MQ6,MQ7,dirac}.

The blackbody electromagnetic energy $W$ in a noncommutavive framework is given by
\ba 
\label{001}
W &=& \frac 12 \int_V u(\nu,T) ~dV,\nonumber\\
\mbox{}&=&\frac12\int_V\left(\varepsilon_0\,\vec{E}\star\vec{E}^\ast+\frac{1}{\mu_0}\,\vec{B}\star\vec{B}^\ast\right)dV\,,
\ea 
where $u(\nu,T)$, $\vec{E}$ and $\vec{B}$ are the electromagnetic energy density, the electric and the magnetic fields, respectively, and the usual scalar product is replaced by the $\star$-product. From the Maxwell's equations in vacuum, it is obtained the following differential equation for the potential vector 
\be
\label{010}
\vec\nabla^2\vec{A}-\frac{1}{c^2}\displaystyle{\frac{\partial^2\vec{A}}{\partial{t^2}}}=0\,,
\ee
where the Lorentz's gauge was applied, namely,$\,\vec\nabla\cdot\vec{A}=0\,$. The equation (\ref{010}) is a wave equation, then the potential vector field $\vec{A}$ is a periodic vector function, with $\vec{A}=\vec{0}\,$ at the boundary of the blackbody cavity.
After a straightforward calculation, Eq.(\ref{010}) is solved and the periodic field vector is obtained as
\be
\label{020}
\vec{A}(\vec{r},t)=\sum_n\,\sum^2_{\lambda=1}\left(\vec{q}\,(\vec{k}_n,\lambda)\,e^{i\,\vec{k}_n\cdot\vec{r}}+\vec{q}^{\;\ast}(\vec{k}_n,\lambda)\,e^{-i\,\vec{k}_n\cdot\vec{r}}\right)  \,, 
\ee
with $\,\vec{k}_n\,$ and $\lambda$ as being, respectively, the radiation propagation direction and the orthogonal directions of polarization. Further, $\,\vec{q}(\vec{k}_n,\lambda)\,$ are independent vectors and orthogonal to $\,\vec{k}_n\,$. As $\vec{E}=-\displaystyle{\frac{\partial{\vec{A}(\vec{r},t)}}{\partial{t}} }$ and $\vec{B}=\vec{\nabla}\times\vec{A}(\vec{r},t)$, Eq.(\ref{001}) is rewritten as
\be
\label{030}
W =2\,V\,\varepsilon_0\sum_n\sum_\lambda\left[\dot{\vec{q}}\star\dot{\vec{q}}^{\;\ast}+\omega_n^2\,(\vec{q}\star\vec{q}^{\;\ast})\right] \,, 
\ee
with $\omega_n^2=k_n^2\,c^2$ and $c=\sqrt{1/\varepsilon_0\,\mu_0}$.
Assuming that
\ba
\label{040}
\vec{q}&=&\frac{1}{2\,\sqrt{2\,V\,\varepsilon_0}}\left(\vec {Q}+i\,\frac{\vec{P}}{\omega_n}\right)\hat{e}(\vec{k}_n,\lambda)\,e^{-i\,\omega_n\,t} \,,\nonumber\\
\vec{q}^{\;\ast}&=&\frac{1}{2\,\sqrt{2\,V\,\varepsilon_0}}\left(\vec{Q}-i\,\frac{\vec{P}}{\omega_n}\right)\hat{e}(\vec{k}_n,\lambda)\,e^{i\,\omega_n\,t} \,,
\ea
the Eq.(\ref{030}) is rewritten as
\be 
\label{050}
W=\frac12\sum_n\sum_\lambda\left(P^2+\omega_n^2\,Q^2+\,\frac{\hbar}{2}\,\omega_n\right) \,,
\ee
with $N=2$. This is the noncommutative version of the Rayleigh-Jeans theorem for the radiation field.

Applying the Boltzmann distribution of energy among an infinite and enumerable number of oscillators confined in a cavity of volume $V$, considering the updated Planck's postulate and Eq.(\ref{050}), one obtains, by analogy to the usual Planck development, the energy of radiation field as being
\be 
\label{060}
W_n =\left(n_\lambda(\omega)+\frac12\right)\,\hbar\,\omega\,,
\ee
where $n_\lambda(\omega)$ is assumed to be the number of photons with frequency $\omega$ and polarization $\lambda$ while, for $n_\lambda(\omega)=0$, there is a non-null ground state energy equal to $\hbar\,\omega/2$. In view of this, the introduction of a noncommutativity, obtained through the replacement of the usual scalar product by the $\star$-product, induces a non-null ground state energy. Note that this approach changes the Planck's radiation law for the blackbody radiation spectrum, given by
\ba 
\label{065}
\rho(\omega,T)&=&\frac{\omega^2}{\pi^2 c^3}\frac{\sum_{n=0}^{\infty}\frac{(n+1/2)\hbar\omega}{kT}e^{-{(n+1/2)\hbar\omega}/{kT}}}{\sum_{n=0}^{\infty}\frac{1}{kT}e^{-{(n+1/2)\hbar\omega}/{kT}}},\nonumber\\
&=& \frac{\omega^2}{\pi^2 c^3}\frac{\sum_{n=0}^{\infty}{(n+1/2)\hbar\omega}e^{-{n\hbar\omega}/{kT}}}{\sum_{n=0}^{\infty}e^{-{n\hbar\omega}/{kT}}},\nonumber\\
&=& \frac{\omega^2}{\pi^2 c^3}\left(\frac{\sum_{n=0}^{\infty}{n\hbar\omega}e^{-{n\hbar\omega}/{kT}}}{\sum_{n=0}^{\infty}e^{-{n\hbar\omega}/{kT}}}+\frac12\hbar\omega\right),\nonumber\\
&=&\frac{\omega^2}{\pi^2c^3}\left(\frac{\hbar\omega}{e^{\hbar\omega/{kT}}-1}+\frac12\hbar\omega\right),
\ea 
where $k$ and $T$ are, respectively, the Boltzmann's constant and temperature. The equation (\ref{065}) is exactly Planck's radiation law\cite{planck4} for the full blackbody energy spectrum.

It is important to notice that the molecules on the blackbody cavity boundary, a conducting surface connected to a thermal reservoir, oscillate and, consequently, emit electromagnetic waves that propagate through the interior of the cavity and then hits the internal surface of the cavity transferring energy to it, inducing the molecules to oscillate and, as a consequence, these molecules emit electromagnetic waves again; this process takes place successively. Those molecules behavior as charged harmonic oscillators and they would still come to thermal equilibrium with
the ambient thermal radiation\cite{planck4}. Indeed, a classical charged harmonic oscillator acquires an average energy equal
to the average energy per normal mode of the surround random classical radiation at the
frequency $\omega_0$ of the oscillator. Therefore, in the limit as the temperature decreases to zero, the
charged harmonic oscillators would be in equilibrium with the random radiation which exists
at absolute zero temperature. This random radiation at the zero absolute is denominated classical
electromagnetic zero-point radiation. Another example of this is the Casimir effect\cite{casimir}, where the force between two uncharged parallel conducting plates depends upon all the radiation surrounding the plates. This force was experimentally measured\cite{casimir,Sparnaay,Lamoreaux,Mohideen,Chan,Bressi} and it was shown that, at absolute zero, the Casimir force does not vanish, but this force was explained quantitatively by the existence of random classical radiation with an energy spectrum $\varepsilon_\omega = (1/2) \hbar\omega$ per normal mode\cite{Boyer1}. Due to this, $N$ is determined as being equal to $2$ in Eq.(\ref{060}) and, consequently, the correct ground state energy, whether for radiation fields or oscillating systems, is obtained.

\section{Conclusion}
The $\star$-product among the phase space coordinates was applied to the process of determining the electromagnetic energy associated with the radiation emitted by the blackbody. With this, the correct relationship that determines the quantization of the harmonic oscillator was obtained, i.e., the ground state energy was obtained and the original Planck's radiation law was extended in order to embrace the ground state energy $(\hbar\,\omega/2)$.

There is another point that should be presented. Consider that the cavity surface has a boundary conditions at the infinity. In this scenario, we assume that the radiation field is free and $N\rightarrow\infty$. Assuming this hypothesis, the NC contribution disappears, $\frac1N \hbar\omega\rightarrow 0$, and Eq.(\ref{060}) reduces to
\be 
\label{068}
W_n=n_\lambda(\omega)\hbar\omega.
\ee
Now, the sum over all propagation direction, Eq.(\ref{050}), does not present a divergence.

At this point, let's put the result, obtained here, in perspective with the canonical quantization\cite{MQ1,MQ3,dirac}. As known from classical mechanics, the Poisson bracket between the phase space coordinates is
\be
\label{070}
\{\,{Q_i}~,~{P_j}\,\}=\delta_{ij}\,,~i,j=1,2,3.
\ee
In this way, the implementation of the canonical quantization process can be performed by applying the following relation:
\be
\label{080}
\{\,{Q_i}~,~{P_j}\,\}=\delta_{ij} ~~\rightarrow~~\left[\,\widehat{Q}_i~,~\widehat{P}_j\right] = i\,\hbar~ \delta_{ij}
\ee
where $\left[~,~\right]$ is the quantum commutator, while $\widehat{Q}_i$ and $\widehat{P}_j$ are quantum operators in the Hilbert space. 

The definition of the $\star$-product between the phase space coordinates, Eq.~(\ref{NovoProduto1}), allows to get the same structure of that defined by the canonical quantization, Eq.~(\ref{080}), which is given by
\be 
\label{090}
Q_i\star Pj - P_j \star Q_i=\imath\hbar\delta_{ij},
\ee
with $N=2$. However, the \textit{status} of phase space coordinates is not lifted to that one carried by the quantum operators undertaking to the Hilbert space. This allows us to infer that aspects assumed to be \emph{quantum} effects can still manifest at the classical level, as long as it is immersed in a noncommutative framework induced by the $\star$-product. It is an alternative representation of quantum mechanics, which may for some purposes
(e.g. to study the classical limit) be more useful than the Hilbert space representation.

In the papers\cite{Boyer1,Boyer2,Boyer3,Boyer4} the Planck's law for blackbody energy spectrum and the ground state energy $(\hbar\,\omega/2)$ where obtained by assuming the Lorentz-invariant electromagnetic radiation at the absolute zero of temperature, without the notion of energy discrete linear in frequency, or Dirac's classical-quantum analogy for the harmonic oscillator, or pure temperature ideas. On the other hand, we get the same result extending the Planck's postulate by using $\star$-product. Further, the modern quantum mechanics also reproduces the same previous result. Due to this, there are, at least, four distinct descriptions of the Planck's law for blackbody energy spectrum and the ground state energy $(\hbar\,\omega/2)$, equivalent when compared at experimental level. This shows that the realm for deciding which theory is correct is not only at the experimental level(e.g. blackbody radiation), but also at the theoretical level. Indeed, over time, modern quantum mechanics started at 1925 has shown to be the theory that best provides a description of the physical properties of nature at the scale of atoms and subatomic particles, far beyond semi-classical limit.

\section{ Acknowledgments}
This work is supported in part by FAPERJ and CNPq, Brazilian
Research Agencies.


\begin{thebibliography}{99}
\bibitem{books} See, for example, R. Eisberg and R. Resnick, Quantum Physics of Atoms,
Molecules, Solids, Nuclei, and Particles, 2nd ed. (John Wiley \& Sons publisher, New York,
1985); J. Leite Lopes, \emph{Estrutura Qu\^antica da Mat\'eria: do \'atomo pr\'e-socr\'atico \`as part\'{\i}culas elementares}, $2^a$ ed., (Editora UFRJ, Rio de Janeiro, 1993); K. S. Krane, Modern Physics, 2nd ed. (Wiley, New York, 1996); J.
R. Taylor, C. D. Zafiratos, and M. A. Dubson, Modern Physics for
Scientists and Engineers, 2nd ed. (Pearson, New York, 2003); S. T.
Thornton and A. Rex, Modern Physics for Scientists and Engineers, 4th
ed. (Brooks/Cole, Cengage Learning, Boston, MA, 2013).


\bibitem{Boyer1} T. H. Boyer, \emph{Derivation of the blackbody radiation spectrum without quantum assumptions}, Phys. Rev. \textbf{182}, 1374–1383 (1969).
\bibitem{Boyer2} Timothy H. Boyer, \emph{Blackbody radiation in classical physics: A historical perspective}, American Journal of Physics \textbf{86}, 495 (2018); revised in November 2, 2018.
\bibitem{Boyer3} T. H. Boyer, \emph{Dirac’s Classical-Quantum Analogy for the Harmonic Oscillator: Classical Aspects in Thermal Radiation Including Zero-Point Radiation}, Am. J. Phys. \textbf{88}, 640 (2020).
\bibitem{Boyer4} T. H. Boyer, \emph{Thermodynamics of the harmonic oscillator: derivation of the Planck blackbody spectrum from pure thermodynamics},  Eur. J. Phys. \textbf{40}, 025101 (2019).

\bibitem{Wigner} Wigner, E., \emph{On the quantum correction for thermodynamic equilibrium}, {Phys. Rev.} {\bf 40}, 749--759 (1932).
\bibitem{Groenewold} H. J. Groenewold, \emph{On the Principles of elementary quantum mechanics}, Physica (Utrecht) \textbf{12}, 405 (1946).
\bibitem{Moyal} J. E. Moyal, \emph{Quantum mechanics as a statistical theory}, Proc. Cambridge Philos. Soc. \textbf{45}, 99 (1949). 
\bibitem{Groot} de Groot S. R. and Suttorp L. G., \emph{Foundation of Electrodynamics}, (Amsterdam: North-Holland, 1972).
\bibitem{Bayen1} Bayen, F., Flato, M., Fronsdal, C., Lichnerowicz, A. and Sternheimer, D., \emph{Deformation theory and quantization. I. Deformations of symplectic structures}, Ann. of Phys., Vol. \textbf{111}, Issue 1, Pages 61-110 (1978).
\bibitem{Bayen2} Bayen, F., Flato, M., Fronsdal, C., Lichnerowicz, A. and Sternheimer, D., \emph{Deformation theory and quantization. II. Physical applications},Ann. of Phys., Vol. \textbf{111}, Issue 1, Pages 111-151 (1978).
\bibitem{lesche1} Lesche, B., \emph{From classical mechanics to Feynman graphs with $\star$-products}, Phys. Rev. \textbf{D29}, 2270 (1984).
\bibitem{lesche2} Lesche, B. and Seligman,T. H.,  \emph{Quantum mechanics in coherent algebras on phase space}, J. Phys. A: Math. Gen. \textbf{19} 91-105 (1986).

\bibitem{Hermann} A. Hermann, \emph{The Genesis of Quantum Theory (1899-1913)}, (translated by C. W. Nash) (MIT
Press, Cambridge MA 1971), pp. 15--17.

%\bibitem{planck1} M. Planck, \emph{Annalen der Physik}, v.\textbf{1}, p.609 (1900). 
%\bibitem{planck2} M. Planck, \emph{\"{U}ber das Gesetz der Energieverteilung im Normalspektrum}, Annalen der Physik, v.\textbf{4}, 553--563 (1901). 
%\bibitem{planck3} M. Planck, \emph{Annalen der Physik}, v.\textbf{4}, p.564 (1901).

\bibitem{MQ1} P.A.M. Dirac, \emph{The fundamental equations of quantum mechanics}, Proceedings of the Royal Society, London, \textbf{A109}, Issue 752, p.642 (1925).
\bibitem{MQ2} Born, M. and Jordan, P., \emph{On Quantum Mechanics}, Zs. f. Phys., v.\textbf{34}, p.858 (1925).
\bibitem{MQ3} P.A.M. Dirac, \emph{On the theory of quantum mechanics}, Proceedings of the Royal Society, London, \textbf{A112}, Issue 762, p.661 (1926).
\bibitem{MQ4} Born, M., Heisenberg, W. and Jordan, P., \emph{Zur Quantenmechanik II}, Zs. f. Phys., v.\textbf{35}, p.557 (1926); , English translation in Ref. 3, paper 15. \href{http://scholar.google.com/scholar_lookup?hl=en&volume=35&publication_year=1926&pages=557-615&journal=\%00null\%00&issue=\%00null\%00&issn=0044-3328&author=M.+Bornauthor=W.+Heisenbergauthor=P.+Jordan&title=Zur+Quantenmechanik+II&pmid=\%00empty\%00&doi=10.1007\%2FBF01379806}{Google Scholar}, Crossref
\bibitem{MQ5} Schr\"{o}dinger, E., \emph{\"{U}ber das Verh\"{a}ltnis der Heisenberg-Born-Jordanschen Quantenmechanik zu der meinem}\emph{Annalen der Physik}, v.\textbf{79}, p.734 (1926).
\bibitem{MQ6} Schr\"{o}dinger, E., \emph{Quantisierung als Eigenwertproblem} (Erste Mitteilung). Ann. Phys., \textbf{384}, 361–376 (1926).
\bibitem{MQ7} Schr\"{o}dinger, E., \emph{Quantisierung als Eigenwertproblem} (Zweite Mitteilung). Ann. Phys., 384, 489–527 (1926).
\bibitem{dirac} P.A.M. Dirac, \emph{Lectures on Quantum Mechanics} (Belfer Graduate School of Science, Yeshiva University, New York, 1964).

\bibitem{planck4} M. Planck, \emph{The Theory of Heat Radiation} (Dover, New York 1959).

\bibitem{casimir} H. B. G. Casimir, \emph{On the attraction between two perfectly conducting plates}, Proc. Ned. Akad. Wetenschap. \textbf{51}, 793-795 (1948).
\bibitem{Sparnaay} M. J. Sparnaay, \emph{Measurement of the attractive forces between flat plates}, Physica (Amsterdam) \textbf{24}, 751-764 (1958).
\bibitem{Lamoreaux} S. K. Lamoreaux, \emph{Demonstration of the Casimir force in the 0.6 to 6 $\mu$ 55m range}, Phys. Rev. Lett. \textbf{78}, 5-8 (1997): 81, 5475-5476 (1998).
\bibitem{Mohideen} U. Mohideen, \emph{Precision measurement of the Casimir force from 0.1 to 0.9 $\mu$m}, Phys. Rev. Lett. \textbf{8}1, 4549-4552 (1998)
\bibitem{Chan} H. B. Chan, V. A. Aksyuk, R. N. Kleinman, D. J. Bishop, and F. Capasso, \emph{Quantum mechanical
actuation of microelectromechanical systems by the Casimir force}, Science \textbf{291}, 1941-1944
(2001).
\bibitem{Bressi} G. Bressi, G. Carugno, R. Onofrio, and G. Ruoso, \emph{Measurement of the Casimir force
between parallel metallic surfaces}, Phys. Rev. Lett. \textbf{88}, 041804(4) (2002).


\end{thebibliography}
\end{document}